# CONTEX-T: Contextual Exploitation of Encrypted Traffic for Device Fingerprinting via Transformer Time-Frequency Analysis

Nazmul Islam, *Graduate Student Member, IEEE,* and Mohammad Zulkernine, *Senior Member, IEEE*

***Abstract*—The rapid expansion of internet of things (IoT) devices has created a pervasive ecosystem where encrypted wireless communications serve as the primary privacy and security protection mechanism. While encryption effectively protects message content, contextual information from packet metadata and statistics inadvertently expose device identities. Various studies have exploited raw packet statistics and their visual representations for device fingerprinting and identification. However, these approaches remain confined to the spatial domain with limited feature representation. Therefore, this paper presents CONTEX-T, a novel framework that exploits device-level information from encrypted traffic metadata using temporal and spectral representation. The experiments show that time-frequency analysis provides new and rich feature representation, revealing a complex and expanding threat landscape that would require robust countermeasures for IoT security management. CONTEX-T first transforms raw packet-length sequences into temporal and spectral representations and then utilizes vision transformers (ViTs) for device identification. We systematically evaluated multiple time-frequency representation techniques and transformer-based models across encrypted traffic samples from various IoT devices. CONTEX-T achieved device classification accuracy exceeding 99% while operating passively on observable contextual metadata. This demonstrates that temporal and spectral signatures persist under strong encryption, highlighting a critical attack surface for IoT network security and management.**

***Index Terms*—Device fingerprinting, IoT security, time-frequency, vision transformer, wireless communication.**

## I. INTRODUCTION

THE exponential growth of internet of things (IoT) devices spans critical domains including smart homes, healthcare monitoring, industrial automation, and urban infrastructure. This growth establishes a ubiquitous digital ecosystem where wireless communications serve as the foundation for device interaction and data exchange [1]. To protect sensitive user information and operational data during the communication, encryption protocols such as Transport Layer Security (TLS), Secure Sockets Layer (SSL), and WiFi Protected Access (WPA) have become standard, forming the primary line of defense against payload-based eavesdropping and content analysis attacks [2]. While these cryptographic protections secure message content, they inadvertently expose a critical vulnerability. Encrypted traffic metadata remains visible to passive observers, enabling covert contextual exploitation of metadata for tasks such as device fingerprinting independent of encryption strength.

For an adversary, device fingerprinting is crucial for cyber intelligence, mission planning, capability analysis and reconnaissance attack that serves as a covert initial step in the broader cyber kill chain [3]. These attacks operate through side-channel analysis of traffic metadata, revealing device types, behavioral patterns, and even enumerate system vulnerabilities. Techniques such as passive packet sniffing can capture all packets traversing a wireless channel using commodity packet-capturing tools such as Wireshark, tcpdump, and Aircrack-ng [4]. Adversaries can intercept these encrypted network packets, extracting critical metadata including packet lengths, source MAC addresses, access points, traffic rates, and temporal patterns without triggering intrusion detection systems or requiring active network participation. Once adversaries compile these fingerprints, they can identify devices and exploitable vulnerabilities for subsequent attack in cyber kill chain. Unlike active scanning techniques (port scanning, service enumeration) that generate anomalous traffic volumes, passive sniffing blends seamlessly into legitimate network activity, providing attackers with persistent, undetected access to intelligence-gathering capabilities.

The vulnerability of encrypted traffic to metadata analysis for device fingerprinting is well established in prior research [5]. Building on this foundation, recent studies have introduced network traffic image analysis as a more effective alternative for device fingerprinting and identification. Visual analysis enhances feature extraction and representation while alleviating the limitations of manual feature engineering and scalability [6], [7], [8], [9]. These methods predominantly convert raw packet streams into spatially organized grayscale images, on which deep learning (DL) models are trained and tested. Such image-based approaches preserve byte-level sequential and spatial locality but fundamentally remain in the amplitude domain. While some temporal modeling techniques have been integrated into these frameworks, they fundamentally treat temporal dynamics (such as periodicity, harmonic structure, and frequency modulations) implicitly. Neural networks must learn to recognize periodicity, polling cycles, heartbeat intervals, and inter-packet timing patterns through data-driven feature extraction. This implicit learning paradigm lacks interpretability, requires large, annotated datasets to capture temporal periodicities, and provides no formal guarantees about which traffic characteristics enable device identification.

Despite the effectiveness of spatial-domain grayscale methods, the time-frequency domain provides enhanced feature

This research is partially supported by the Canada Research Chairs Program and the Natural Sciences and Engineering Research Council of Canada.

Nazmul Islam and Mohammad Zulkernine are with the School of Computing, Queen's University, Kingston, ON K7L 3N6, Canada (e-mail: {nazmul.islam, mz}@queensu.ca).

representation and interpretability, as established in prior work across domains such as human activity recognition [10], emotion recognition [11], intrusion detection [12], [13], internet traffic analysis [14], and healthcare [15]. However, systematic applications and evaluations of time-frequency domain analysis for IoT device fingerprinting are limited in the literature. Furthermore, the integration of vision transformers (ViTs) with time-frequency analysis introduces a new architectural paradigm for IoT fingerprinting. Unlike CNNs that capture local spatial correlations through convolutional kernels, ViTs leverage global self-attention mechanisms to model long-range dependencies across time-frequency representations. This enables the capture of distant time-frequency relationships (such as harmonics separated by multiple frequency bins and periodic bursts separated by time lags) [16].

Therefore, this paper presents CONTEX-T, a novel framework that exploits contextual metadata in encrypted wireless IoT traffic through time-frequency analysis with rich feature representation. CONTEX-T systematically decomposes raw packet-length sequences into temporal and spectral representation using short-time Fourier transform (STFT) and continuous wavelet transform (CWT), extracting explicit information-theoretic features including time-frequency peaks, harmonic decomposition, instantaneous frequency, phase coherence, and wavelet variance. We evaluate state-of-the-art ViT architectures on these representations, achieving superior device classification accuracy across encrypted traffic samples while maintaining complete passivity and undetectability. Through statistical validation using 95% bootstrap confidence intervals, we demonstrate that device-specific communication behaviors persist as temporal and spectral signatures even under strong encryption. Importantly, the proposed approach reveals a new threat landscape of time-frequency analysis with rich feature representation that adversaries may exploit for device fingerprinting. The primary contributions of this paper are:

- *Time-Frequency Domain:* Introduction of time-frequency representation (STFT/CWT) of encrypted packet-length sequences that provide device specific time-frequency features instead of implicit learned representation in IoT.
- E*mpirical Validation of Time-Frequency Representations*: Unsupervised and semi-supervised evaluation providing model-independent empirical evidence that time-frequency decomposition exposes device-discriminative structure that remains latent in spatial (grayscale) representations.
- *Transformer Architecture Benchmarking*: Systematic evaluation of ViT models with statistical validation via 95% bootstrap confidence intervals. This determines the optimal combinations of time-frequency resolution, segment length, and overlap for best performance.
- *Generalization and Cross-Configuration Evaluation*: Robustness analysis assessing model performance across held-out out-of-distribution (OOD) evaluation set. Identifying configurations that maintain high accuracy with these datasets that reflect variable observation windows and establish practical limitations.

The remainder of this paper is organized as follows: Section II surveys related work on traffic analysis attacks and IoT device fingerprinting. Section III provides the methodology for time-frequency decomposition and ViT models. Section IV presents the adversarial model and CONTEX-T framework. Section V reports experimental setup, evaluation, and cross-configuration generalization. Section VI concludes the paper.

## II. RELATED WORK

The related studies are divided into two subsections, DL-based IoT fingerprinting from raw data and from image representations.

### *A. DL-Based IoT Fingerprinting from Raw Traffic*

The earliest approaches to IoT device identification relied on direct analysis of packet-level features extracted from raw traffic. These methods leverage statistical properties, temporal patterns, and distributional characteristics of raw metadata to construct device fingerprints. Protocol-aware statistical features were used in [17] that rely on multiple flow attributes (volume, duration, and sleep time) together with protocol-specific signaling patterns (DNS queries, NTP timestamps, and cipher suites) to classify IoT devices. However, the approach requires extensive protocol parsing and domain-specific expertise. The frequency distribution modeling of ByteIoT [18] eliminates protocol dependencies by distilling identification to packet-length frequency distributions processed via Hellinger distance-based k-nearest neighbor classification with DBSCAN clustering. IoTDevID [19] advances cross-dataset generalization through systematic elimination of session-specific information. The study applied decision trees with MAC-based aggregation to solve the transfer problem for devices sharing gateway addresses. Time-series dictionary approach (ROCKET + HYDRA) in [20] streamlines feature extraction by applying 10,000 random convolutional kernels over multivariate packet-length sequences to capture multi-scale temporal patterns and carried out logistic regression classification with minimal computational overhead. MetaRockETC [21] extends this paradigm through meta-learning, leveraging model-agnostic meta-learning (MAML) over multivariate time-series representations (original, uplink/downlink, cumulative sums).

Despite these advances, all five methods rely on explicit feature extraction or transformation pipelines that either require domain expertise (statistical feature selection), extensive hyperparameter tuning (kernel transformations, distance metrics), or multi-stage processing (packet-level classification followed by aggregation), limiting end-to-end optimization and potentially leaving discriminative patterns unexploited [5]. This limitation motivates a paradigm shift toward spatial-domain representations, where raw packet sequences are reshaped into 2D grayscale images for DL analysis discussed in the following section.

### *B. DL-Based IoT Fingerprinting from Image Representations*

Techniques in [6], [7], [8], [9] convert N-byte sessions into fixed-size grayscale images, where each byte $b_i \in$ directly maps to pixel intensity $I(x, y) = b_i$. These techniques employed diverse DL approaches to address manual feature engineering and scalability challenges. DMRMTT [6] combines a deep convolutional maxout network with multiple time-series transformers to extract spatial and temporal features from 28×28 grayscale fingerprints. Authors in [7] proposed a depth wise

TABLE I

SUMMARY AND COMPARISON OF EXISTING LITERATURE ON DEVICE IDENTIFICATION USING IMAGE REPRESENTATIONS

| Reference | Feature Domain | Traffic Pattern Representation | Traffic Pattern Recognition | Information preserved | Interpretability | Metadata utilized | Robustness to Jitter | DL Architecture |
|---|---|---|---|---|---|---|---|---|
| **DMRMTT [6]** | Spatial | Grayscale Image (Spatial Byte Patterns) | Implicit CNN Learning (Black-box) | Spatial Locality | Low (Texture) | - | Low (Order-sensitive) | Maxout + MTT (CNN-based) |
| **Feng et al. [7]** | | | | | | Multiple* | Medium (GRU Smoothing) | DSC + GRU (CNN-based) |
| **SLIoTDI [8]** | | | | | | - | Low (Single-session) | LeNet + FGSM (CNN-based) |
| **IoT-SCNet [9]** | | | | Spatial Locality (Multi-level) | Medium (Feature Fusion) | Multiple* | Medium (Multi-level) | ResNet-50 (CNN-based) |
| **Proposed Work** | Time-frequency | Spectrograms and Scalograms | Explicit Spectral Features | Multi-resolution Time-frequency, Amplitude, Phase | High (Spectral peaks, Bandwidth) | Only Packet Length | High (Magnitude-invariant) | ViT (Global Self-attention) |

* Includes few or more additional metadata such as packet size, direction (in/out), inter-arrival Time, RTT timing, protocol type, TLS/Cipher.

separable convolution (DSC) and GRU-based framework using triplet loss for automatic fingerprint extraction from traffic image sequences. SLIoTDI [8] introduced a modular, reusable feature extractor via adversarial training (FGSM) and a joint loss function combining paired, adversarial, and cooperative losses, enabling scalability to unseen devices without retraining. IoT-SCNet [9] leveraged semi-supervised contrastive learning with traffic-specific augmentation to reduce labeled data dependency.

These approaches preserve spatial locality and sequential byte information, but it remains fundamentally limited to amplitude encoding. Additionally, they treat temporal information implicitly where the networks must learn to recognize periodicity, polling cycles, and inter-packet timing patterns through feature extraction rather than explicit frequency decomposition.

### C. Proposed Time-frequency Representations

The fundamental distinction for feature representation between CONTEX-T and existing grayscale image-based approaches lies not only in raw feature extraction, but also in the representational domain and explicit temporal-frequency decomposition. While paper [6], [7], [8], [9] operate in the spatial domain (directly mapping raw packet sequences to pixel intensities), CONTEX-T shifts the entire analytical paradigm to the time-frequency domain, enabling explicit extraction of periodic device behaviors and time-frequency signatures that remain hidden in spatial representations. Table I provides comparison of this work with the existing literature for packet representation.

CONTEX-T applies STFT and CWT directly to packet-length sequences, producing temporal and spectral representations (spectrograms and scalograms) where time-frequency peaks reveal device-specific polling cycles and communication patterns invisible to spatial grayscale encodings. These representations offer key advantages: (1) Information preservation through explicit temporal-frequency decomposition that exposes device-specific periodicities, bursts, and modulation patterns beyond first-order spatial statistics; (2) Tunable resolution via adaptive STFT window lengths and CWT scales, balancing temporal/frequency precision; (3) Robustness through STFT magnitude translation-invariance ($|S(\omega)|^2$) and CWT scale-invariance, mitigating packet jitter/reordering; and (4) Multi-resolution analysis for optimization. Empirical validation of these time-frequency representations is presented in Section V.C.

## III. METHODOLOGY

This section outlines the methodological framework used in the research. We describe the techniques employed to generate time-frequency representations and detail the ViT models.

### A. Time-frequency Decomposition Techniques

To extract discriminative features suitable for DL classification, we transform one-dimensional raw packet sequences into two-dimensional time-frequency representations. This representation simultaneously captures two complementary information dimensions: temporal evolution and frequency content. Time-frequency analysis reveals which frequency components (corresponding to transmission rate patterns) are active at which temporal intervals, enabling neural networks to learn device-specific behavioral signatures that would remain hidden in purely temporal or purely frequency-domain representations.

Let the packet length sequence be denoted as $p = [p_1, p_2, \dots, p_{L_{seg}}]$, where $p_n \in \mathbb{R}^+$ represents the length in bytes of the $n$-th packet and $L_{seg}$ is the segment length. The objective is to compute a representation $S(t, f)$ that encodes the time-frequency energy distribution at each temporal location, enabling the extraction of localized time-frequency features for classification. The fundamental trade-off in this analysis, formalized by the Heisenberg-Gabor uncertainty principle [22], states that simultaneous perfect localization in both time and frequency is impossible:

$$\Delta t \cdot \Delta f \geq \frac{1}{4\pi}. \quad (1)$$

This principle guides the choice between STFT (fixed resolution) and CWT (adaptive resolution), as both methods represent different strategies for resolving this trade-off.

1) **Short-Time Fourier Transform (STFT)**

The STFT provides a time-localized frequency decomposition by applying the Fourier transform to windowed segments [23]. Algorithm 1 shows STFT-based spectrogram generation. For a packet sequence $p[n]$, the STFT is:

$$\mathrm{STFT}_m^k = \sum_{n=0}^{R-1} p[m \cdot h + n] \cdot w[n] \cdot e^{-\frac{j2\pi kn}{K}}, \quad (2)$$

where, $m$ is the time-frame index (indicating which window

**Algorithm 1:** STFT-based Spectrogram Generation

**Input**: Packet segment $\widetilde{\mathbf{p}}_i \in \mathbb{R}^{L_{\text{seg}}}$, frequency resolution $R$, stride $f_{\text{stride}}$
**Output**: STFT spectrogram $S_{\text{STFT}} \in \mathbb{R}^{T\times(R/2+1)}$
1. **Initialize:**
• Window length: $L \leftarrow R$
• FFT points: $K \leftarrow R$
• Hop size: $h \leftarrow \lfloor L \cdot f_{\text{stride}} \rfloor$
• Number of frequency bins: $F \leftarrow K/2+1$
• Hann: $w[n] \leftarrow 0.5(1-\cos(2\pi n/(R-1)))$ for $n=0,\dots,R-1$
• Sampling rate: $f_s \leftarrow 1.0$ cycle/packet
• Frequency axis: $f[k] \leftarrow (k\cdot f_s)/K = k/R$ for $k=0,1,\dots,F-1$
• Number of frames: $T \leftarrow \lfloor (L_{\text{seg}}-R)/h \rfloor + 1$
**2. Compute STFT:**
• for $m=0$ to $T-1$ do
• for $k=0$ to $F-1$ do
• $\text{STFT}[m,k] \leftarrow \sum_{n=0}^{R-1} \widetilde{\mathbf{p}}_i[m\cdot h+n]\cdot w[n]\cdot e^{-j2\pi kn/K}$
• end for
• end for
**3. Convert to Spectrogram:**
• for $m=0$ to $T-1$ do
• for $k=0$ to $F-1$ do
• $S_{\text{STFT}}[m,k] \leftarrow 10\log_{10}(|\text{STFT}[m,k]|^2 + 10^{-12})$
• end for
• end for
**4. Return $S_{\text{STFT}} \in \mathbb{R}^{T\times(R/2+1)}$**

position), $h$ is the hop size (stride between consecutive windows), $w[n]$ is the window function of length $R$, $k$ is the frequency bin index, $K$ is the FFT size determined by the window length parameter, and $j=\sqrt{-1}$. We employ the Hann window for its favorable spectral leakage properties that minimize sidelobe leakage [24]:

$$w[n] = 0.5\left(1-\cos\left(\frac{2\pi n}{R-1}\right)\right), n=0,1,\dots,R-1. \quad (3)$$

The STFT coefficients $\text{STFT}_m^k$ are complex-valued. The power spectrum is computed as the magnitude-squared, then converted to decibel scale for perceptual relevance:

$$S_{\text{STFT}}[m,k] = 10\log_{10}(|\text{STFT}_m^k|^2+\varepsilon), \quad (4)$$

where $\varepsilon = 10^{-12}$ prevents instability from logarithm of zero.

In STFT, the frequency axis has uniformly spaced bins:

$$f_k = \frac{k\cdot f_s}{K}, \quad (5)$$

where $f_s = 1.0$ cycle/packet (normalized sampling rate). The frequency resolution is:

$$\Delta f = \frac{f_s}{R}, \quad (6)$$

where $R$ is the window length parameter. In the implementation, the spectral resolution parameter $R \in \{16,32,64\}$ determines the window length (with $K=R$), yielding three spectral resolution levels: $R=16$ for coarse frequency resolution $\Delta f = 1/16$ and fine temporal resolution $\Delta t = 16$ packets; $R=32$ for medium resolution in both domains; and $R=64$ for fine frequency resolution $\Delta f = 1/64$ and coarse temporal resolution $\Delta t = 64$ packets.

The limitation of STFT is its fixed time-frequency resolution. A window of length $R$ determines both temporal and frequency localization with inverse relationship which cannot be separately optimized. Increasing $R$ improves frequency resolution but worsens temporal resolution, preventing accurate simultaneous localization of transient events and narrow

**Algorithm 2:** CWT-based Scalogram Generation

**Input**: Packet segment $\widetilde{\mathbf{p}}_i \in \mathbb{R}^{L_{\text{seg}}}$, frequency resolution $R$,
**Output**: CWT scalogram $S_{\text{CWT}} \in \mathbb{R}^{R\times L_{\text{seg}}}$
1. **Initialize:**
• Number of scales: $N_s \leftarrow R$
• Morlet wavelet center frequency: $f_c$
• Sampling rate: $f_s \leftarrow 1.0$ cycle/packet
• Frequency bounds:
• $f_{\min} \leftarrow f_s/(2R) = 1/(2R)$ (minimum resolvable frequency)
• $f_{\max} \leftarrow f_s/2 = 0.5$ (Nyquist frequency)
• Numerical stability constant: $\varepsilon \leftarrow 10^{-12}$
**2. Generate Logarithmically-spaced Frequency and Scale Arrays:**
• for $i=0$ to $N_s-1$ do
• $f[i] \leftarrow f_{\min}\cdot(\frac{f_{\max}}{f_{\min}})^{i/(N_s-1)}$
• $s[i] \leftarrow \frac{f_c\cdot f_s}{f[i]} = \frac{0.8125}{f[i]}$
• end for
• *Note*: This creates R scales ordered from largest (index j=0, lowest frequency) to smallest (index j=R-1, highest frequency)
3. **Compute CWT:**
• for $j=0$ to $N_s-1$ do
• for $n=0$ to $L_{\text{seg}}-1$ do
• Compute: $W[j,n] \leftarrow \frac{1}{\sqrt{s[j]}}\sum_{n'=0}^{L_{\text{seg}}-1} \widetilde{\mathbf{p}}_i[n']\cdot\bar{\psi}(\frac{n'-n}{s[j]})$
• end for
• end for
4. **Convert to Scalogram:**
• for $j=0$ to $N_s-1$ do
• for $n=0$ to $L_{\text{seg}}-1$ do
• $S_{\text{CWT}}[j,n] \leftarrow 10\log_{10}(|W[j,n]|^2 + 10^{-12})$
• end for
• end for
5. **Return $S_{\text{CWT}} \in \mathbb{R}^{R\times L_{\text{seg}}}$**

frequency components. This uniform trade-off applies identically across all frequencies, making STFT particularly effective for stationary signals but less suitable for highly non-stationary traffic patterns.

2) **Continuous Wavelet Transform (CWT)**

CWT circumvents the STFT resolution limitation through scale-dependent analysis. Algorithm 2 shows CWT-based scalogram generation. CWT uses scaled and translated copies of a single mother wavelet function [25], [26]:

$$W(a,b) = \frac{1}{\sqrt{a}}\int_{-\infty}^{\infty} p(t)\cdot\bar{\psi}\left(\frac{t-b}{a}\right)dt, \quad (7)$$

where, $a \in \mathbb{R}^+$ is the scale parameter (small scales correspond to high frequencies, large scales correspond to low frequencies), $b \in \mathbb{R}$ is the translation parameter (temporal localization), $\psi(t)$ is the mother wavelet function and $\bar{\psi}$ denotes the complex conjugate.

For implementation with packet sequences, we define the resolution $R \in \{16,32,64\}$ as the number of scales at which the CWT is evaluated. The scales $\{s_j\}_{j=1}^R$ are derived from $R$ log-spaced frequencies $\{f_j\}_{j=1}^R$ between $f_{\min}$ and $f_{\max}$ via:

$$s_j = \frac{f_c\cdot f_s}{f_j}, j=1,2,\dots,R, \quad (8)$$

where $f_c$ is the center frequency of the Morlet wavelet and $f_s = 1.0$ is the sampling rate (cycles/packet). CWT's defining characteristic is its adaptive resolution. Each scale corresponds to a pseudo-frequency through equation (8). This mapping creates multi-resolution analysis. High frequencies (small scales) create fine temporal resolution, coarse frequency resolution and low frequencies (large scales) create coarse temporal resolution, fine frequency resolution.

The CWT is computed as:

$$W[j,n] = \frac{1}{\sqrt{s_j}} \sum_{n'=0}^{N-1} p[n'] \cdot \bar{\psi}\left(\frac{n'-n}{s_j}\right), j = 1,2,\dots,R, \quad (9)$$

where the sum is over the packet indices. We employ the complex Morlet wavelet with a center frequency of $f_c$ cycles/sample (normalized frequency). This is a standard implementation used in widely deployed signal processing libraries for computational consistency. The Morlet wavelet combines sinusoidal oscillation with a Gaussian envelope, providing excellent joint localization in both time and frequency domains. This makes it well-suited for detecting transient packet rate changes superimposed on sustained background traffic patterns [25], [27]. The CWT power scalogram is converted to decibel:

$$S_{\text{CWT}}[j,n] = 10\log_{10}(|W[j,n]|^2 + \varepsilon). \quad (10)$$

Formally, the time-frequency resolution cells have area inversely proportional to frequency:

$$\Delta t(f) \cdot \Delta f(f) \propto \frac{1}{f}. \quad (11)$$

This means CWT achieves better frequency resolution at low frequencies (useful for sustained traffic patterns) and better time resolution at high frequencies (useful for burst detection). Additionally, CWT provides multiresolution analysis where resolution varies with frequency as opposed to STFT which maintains constant time-frequency resolution across all frequencies [28], [29].We use the same three resolution levels $R \in \{16,32,64\}$ for CWT as for STFT, where $R$ controls the number of frequency points in the analyzed range. This enables fair comparison of device classification accuracy across both transformation methods.

### *B. Vision Transformer Models*

This research employs three state-of-the-art vision transformer (ViT) architectures for classifying IoT devices based on their representations. DeiT-Base (data-efficient image transformer - base) [30], ViT-Small (vision transformer - small) [31] and EfficientViT-B1 (efficient vision transformer - base variant 1) [32]. The ViT processes input images through a sequence of operations that transform 2D spatial data into a 1D sequence of embeddings suitable for transformer processing [33], [34]. It partitions input representations $X \in \mathbb{R}^{H \times W \times C}$ into non-overlapping patches of size $P \times P$ pixels. For standardized input size $H = W = 224$ and $C = 3$ (RGB), the image yields $N = \frac{H \cdot W}{P^2} = \frac{224 \times 224}{16^2} = 196$ patches. Each patch $x_i \in \mathbb{R}^{P^2 \cdot C}$ is linear and projected to a d-dimensional embedding:

$$z_i = E \cdot \text{vec}(x_i) + b_e, \quad (12)$$

where $E \in \mathbb{R}^{d \times P^2 C}$ is the patch embedding matrix, vec $(x_i)$ is the vectorized patch, and $b_e \in \mathbb{R}^d$ is the embedding bias. Learnable position embeddings $p_i \in \mathbb{R}^d$ encode spatial location information:

$$z_i^{\text{pos}} = z_i + p_i. \quad (13)$$

A learnable class token $z_{\text{cls}} \in \mathbb{R}^d$ is prepended to the sequence of $N + 1 = 197$ tokens:

$$Z^{(0)} = \left[z_{\text{cls}}; z_1^{\text{pos}}; \dots; z_N^{\text{pos}}\right] \in \mathbb{R}^{(N+1) \times d}. \quad (14)$$

This sequence is processed through $L$ transformer encoder layers. Each layer $l$ applies multi-head self-attention (MSA) with residual connection:

$$A^{(l)} = \text{MSA}\left(\text{LN}\left(Z^{(l-1)}\right)\right), \quad (15)$$

$$Z'^{(l)} = Z^{(l-1)} + A^{(l)}, \quad (16)$$

followed by a feed-forward network (FFN):

$$F^{(l)} = \text{FFN}\left(\text{LN}\left(Z'^{(l)}\right)\right), \quad (17)$$

$$Z^{(l)} = Z'^{(l)} + F^{(l)}, \quad (18)$$

where $l \in \{0,1,\dots,L\}$, with $l = 0$ representing the input token sequence and $l = L$ representing the final transformed representation. LN denotes layer normalization. Multi-head attention with $H$ heads computes:

$$\text{Attention}(Q,K,V) = \text{softmax}\left(\frac{QK^T}{\sqrt{d_k}}\right)V, \quad (19)$$

with $d_k = d/H$ per head. After $L$ layers, the class token embedding $z_{\text{cls}}^{(L)}$ is extracted and passes a classification head:

$$\hat{y} = \text{softmax}\left(W_{\text{head}} z_{\text{cls}}^{(L)} + b_{\text{head}}\right), \quad (20)$$

$$d_{\text{pred}} = \arg\max_c \hat{y}_c, \quad (21)$$

where $W_{\text{head}} \in \mathbb{R}^{|D| \times d}$ and $|D| = 14$ device classes.

## IV. SYSTEM DESIGN

This section presents the adversarial threat model considered in this work and the overall system architecture. This study adopts the downstream device classification task as prior work [6], [7], [8], [9], and introduces new time-frequency representations of packet-length sequences as input data. While framed from an adversarial exploitation perspective, the core technical contribution remains improved representation learning for the established device fingerprinting problem.

### *A. Adversarial Threat Model*

The adversary, denoted $\mathcal{A}$, is modeled as a passive eavesdropper operating under the honest-but-curious paradigm. The adversary captures packet metadata via wireless sniffing but does not modify, inject, or probe network traffic. We assume the adversary has access to a pretrained fingerprinting model $f_\theta$ obtained from a public labeled dataset or from an offline profiling phase and uses this model for passive inference on newly observed traffic. The adversary is assumed to operate within standard 802.11 Wi-Fi communication range with commodity hardware, consistent with realistic physical proximity attacks. Regarding channel conditions, commodity capture tools such as Wireshark reliably recover packet-length metadata under typical Wi-Fi contention, the effect of packet loss or high channel-interference environments is out of scope.

The adversary is subject to several constraints that define the threat model's scope. First, the payload confidentiality constraint ensures that encrypted application-layer content cannot be inspected, restricting the adversary's inference to observable metadata, specifically packet lengths. Second, the strict passivity constraint prohibits the adversary from actively interacting with the network, including packet injections, timing manipulation, or protocol-level probing. The adversary can only observe traffic and does not have the capability to modify network conditions or inject false data. Third, the temporal boundedness constraint reflects that the pretrained model is fixed at deployment time, having been trained on a specific set of device types and firmware versions; consequently, model accuracy may degrade under distribution shift caused by firmware updates, behavioral changes, or the introduction of previously unseen device types. No adaptive retraining is possible during deployment, which may degrade classification accuracy under non-stationary traffic conditions where packet size distributions evolve over time. This deployment scenario is evaluated in Section V.E, where pretrained models are tested on unseen OOD observation windows with different segment lengths and overlap configurations. Fourth, the limited protocol visibility constraint prevents the adversary from achieving full recovery of higher-layer semantics or infer fine-grained behavioral details, limiting the threat scope to device identification. Additionally, the adversary focuses on device classification given per-device packet-length time series, not on the low-level demultiplexing step. This scoping is consistent with prior work in traffic analysis [6], [7], [8], [9] where the classification task using metadata is considered an independent research question, while demultiplexing is considered a separate problem.

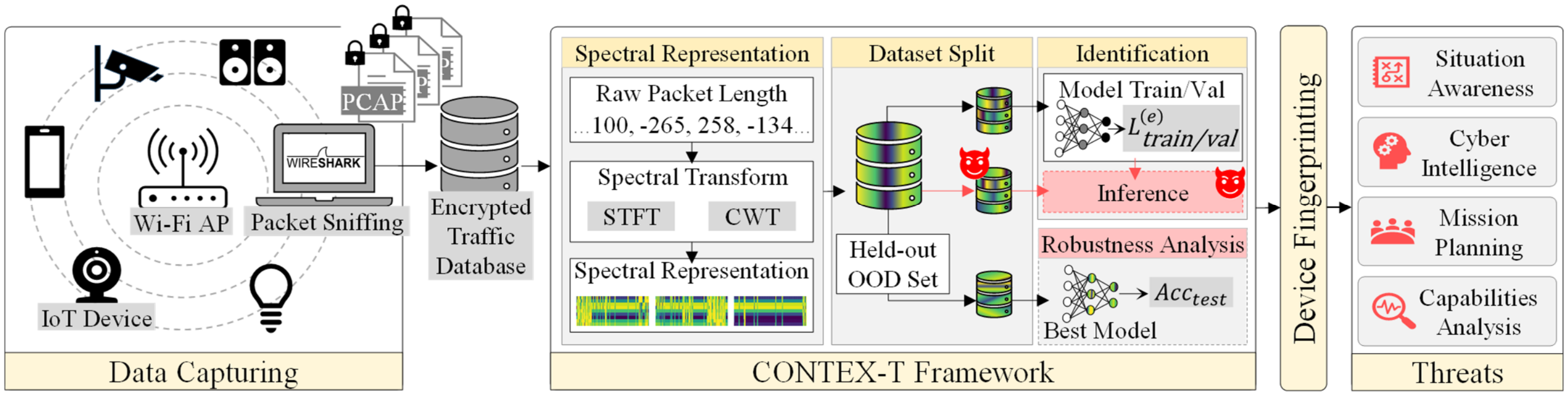

**Fig. 1.** Overview of the proposed framework.

*Adversarial Objectives and Success Metrics:* An adversarial identification attack is considered successful if the test-set accuracy exceeds $\mathrm{Acc}_{\mathrm{test}} \geq 90\%$ with a bootstrap confidence interval width $\Delta_{\mathrm{CI}} < 5\%$, where:

$$\mathrm{Acc}_{\mathrm{test}} = \left(\frac{1}{\mathrm{N}_{\mathrm{test}}}\sum_{\mathrm{i}=1}^{\mathrm{N}_{\mathrm{test}}} \mathbf{1}[\hat{\mathrm{y}}_{\mathrm{i}} = \mathrm{y}_{\mathrm{i}}]\right) \times 100\%, \tag{22}$$

with $\mathrm{N}_{\mathrm{test}}$ = test samples across 14 devices. This significantly exceeds the random-guessing baseline of $1/|\mathrm{D}| \approx 7.14\%$ for 14-way classification. The confidence interval requirement ensures that the success claim is statistically robust and not due to random variation in the test set.

Beyond the primary success criterion, the adversary pursues several secondary objectives to ensure robust and reliable device identification across operational conditions. Cross-configuration robustness on OOD measures whether the attack maintains consistent performance across the configuration setups. This is quantified by minimizing the accuracy variance:

$$\min_{c_1,c_2\in\mathcal{C}} |\mathrm{Acc}_{\mathrm{test}}(c_1) - \mathrm{Acc}_{\mathrm{test}}(c_2)| \approx 0. \tag{23}$$

This ensures that the attack is not sensitive to the specific decomposition parameters and remains effective regardless of which transformation or resolution is employed. Balanced per-class performance prevents classification bias toward frequent devices, which adversaries must overcome for reliable identification across the device distribution. This is measured using the weighted F1-score:

$$\text{Weighted } F_1 = \sum_{d\in D} w_d F_1^{(d)} \geq 0.90, \tag{24}$$

where $w_d = n_d/N_{test}$ is the normalized support weight and $F_1^{(d)}$ is the F1-score for device $d$. A high weighted F1-score indicates that the adversary can accurately identify all device types. Invariance to segmentation parameters captures whether classification accuracy remains stable across different segment lengths $L_{\mathrm{seg}} \in \{100, 500\}$, and overlap $p_{\mathrm{overlap}} \in \{0\%, 50\%\}$:

$$\mathrm{Acc}_{\mathrm{test}}(L_{\mathrm{seg}}, p_{\mathrm{overlap}}) \approx \text{constant}. \tag{25}$$

This ensures that the attack generalizes regardless of how packet sequences are segmented for analysis.

*B. System Design*

Fig. 1 illustrates the overview of the CONTEX-T framework. Building on prior work [6], [7], [8], [9], this study introduces downstream device classification task based on time-frequency analysis of packet-length sequences using ViTs. Training and validation are used for model development, whereas inferencing is performed under the assumed threat model. This section presents the system design of CONTEX-T.

**1) Heterogeneous IoT Traffic Representation**

We consider a heterogeneous internet of things (IoT) network composed of $|D| = 14$ distinct device types communicating over encrypted channels. While application-layer payloads are protected by cryptographic mechanisms, packet-level metadata, specifically packet lengths remain observable in plaintext within network layers to support routing and link-layer operations. The traffic of each device $d \in D$ is modeled as a temporal sequence of packet lengths:

$$p_d = [p_{d,1}, p_{d,2}, \dots, p_{d,T_d}], p_{d,i} \in \mathbb{Z}^+, \tag{26}$$

where $p_{d,i}$ denotes the length in bytes of the $i$-th packet emitted by device $d$. For subsequent analysis, each traffic stream is partitioned into windows of fixed size $L_{seg}$ packet segments:

$$p_d^{(r)} = [p_{d,r,1}, p_{d,r,2}, \dots, p_{d,r,L_{seg}}] \in \mathbb{Z}_+^{L_{seg}}, \tag{27}$$

where $r$ indexes the window position within the device's trace. The segmentation process incorporates partial overlap between

TABLE II

SUMMARY OF DL AND TRAINING CONFIGURATION

| DL Configuration | Specification |
|---|---|
| DL Architecture: | 1. DeiT-Base<br>• Parameters (~86M), Dimension (768)<br>• Transform Layer (12), Attention Head (12)<br>• Transfer: ImageNet + knowledge Distillation |
| | 2. ViT Small<br>• Parameters (~22M), Dimension (384)<br>• Transform Layer (12), Attention Head (6)<br>• Transfer : ImageNet |
| | 3. EfficientViT-B1<br>• Parameters (~7M) Dimension (256)<br>• Transform Layer (4), Attention Head (4)<br>• Transfer: ImageNet, Efficient Attention |
| Input Specification: | PNG, 224x224 pixels, 3 channel (RGB) |
| Learning Rate: | $1 \times 10^{-4}$, Scheduler (OneCycleLR) |
| Optimizer: | AdamW, Adam Parameter ($\beta_1$, $\beta_2$) (0.9, 0.999) |
| Weight Decay: | L2 Regularization (0.05) |
| Gradient Clipping: | Max Norm (1.0) |
| Loss Function: | CrossEntropyLoss |

consecutive windows, parameterized by $L_{seg} \in \{100,500\}$ and $p_{overlap} \in \{0\%, 50\%\}$.

To generate multiple training examples from raw packet sequences while preserving local temporal patterns, we apply sliding window segmentation. Given a full packet sequence $p$, we extract overlapping segments of length $L_{seg}$:

$$p_i = \left[p[i \cdot s_{stride}], p[i \cdot s_{stride} + 1], \dots, p[i \cdot s_{stride} + L_{seg} - 1]\right], \quad (28)$$

where $i = 0,1, \dots, \lfloor (N - L_{seg})/s_{stride} \rfloor$ is the segment index. The stride is determined by the desired overlap percentage:

$$s_{stride} = L_{seg} \cdot \left(1 - p_{overlap}\right), \quad (29)$$

where $p_{overlap} \in \{0\%, 50\%\}$ specifies 0% (non-overlapping) or 50% (50% overlap) configurations. The relationship between stride and overlap is: $stride_{fraction} = 1 - p_{overlap}$. For example, with $L_{seg} = 100$ packets and $p_{overlap} = 0.5$ (50% overlap): $s_{stride} = 100 \cdot (1 - 0.5) = 50$ packets, meaning consecutive segments begin 50 packets apart, creating 50% overlap between adjacent segments for $L_{seg} = 100$.

This windowing strategy achieves a balance between temporal resolution and local stationarity, a property required for subsequent time-frequency decomposition. Overlapping windows create smoother temporal transitions in the representation sequence, capturing behavioral signatures that might be lost at segment boundaries. Additionally, overlapping generates more training samples without requiring additional data collection, which is critical for DL models that benefit from larger training sets while mitigating overfitting risks.

**2) Feature Extraction and Classification Pipeline**

Each traffic segment $p_d^{(r)}$ undergoes a five-stage feature extraction and classification pipeline. Stage 1 is preprocessing and mean-centering, defined as:

$$\tilde{p}_d^{(r)} = p_d^{(r)} - \mu_d, \text{ where } \mu_d = \frac{1}{L_{seg}} \sum_{i=1}^{L_{seg}} p_{d,r,i}. \quad (30)$$

This step removes device-specific bias in packet size distributions while preserving relative variations. The second stage applies a time-frequency transformation, denoted as:

$$T(\tilde{p}_d^{(r)}) = \begin{cases} \text{STFT}\left(\tilde{p}_d^{(r)}; R\right), & \text{if using STFT} \\ \text{CWT}\left(\tilde{p}_d^{(r)}; R\right), & \text{if using CWT} \end{cases}, \quad (31)$$

where the parameter $R \in \{16,32,64\}$ controls the resolution. The choice of transformation (STFT vs CWT) is fixed for each experimental configuration. Stage 3 performs percentile normalization and estimate 5th/95th percentile intensity values from training samples and use these fixed statistics to clip and rescale the dataset, concentrating dynamic range on typical time-frequency patterns. Stage 4 encodes the sample into a visual format by resizing it to 224×224 pixels (to match the input size of the ViT models) and mapping it to representation:

$$X_{d,r} = \text{ResizeRGB}\left(\hat{S}_d^{(r)}, 224 \times 224\right). \quad (32)$$

Finally, the fifth stage performs device classification using a pretrained ViT model $f_\theta(\cdot)$, producing a probability vector:

$$y_{d,r} = f_\theta(X_{d,r}) \in [0,1]^{|D|}, \sum_{c=1}^{|D|} y_{d,r,c} = 1, \quad (33)$$

from which the predicted device label is obtained as:

$$d_{\text{pred}}^{(r)} = \arg\max_{c \in D} y_{d,r,c}. \quad (34)$$

**3) Training Configuration**

Summary of DL and training configuration is given in Table II. In the training pipeline, the data is first partitioned into training, validation, and test sets. Normalization statistics are computed and data augmentation are done only on training set, and the model is optimized according to the specified learning rate schedule. Training proceeds in epochs with early stopping. This structured approach ensures that model development is isolated from evaluation data, preventing data leakage and enabling honest assessment of model performance.

*Dataset Partitioning and Normalization:* From the available representations generated from the raw packet lengths, training, validation and test set are created via a randomized device-specific split that would be detailed in section V.A. Reproducibility is ensured through seeded random number generation which guarantees deterministic and reproducible partitioning across runs and environments.

Instead of using standard ImageNet normalization, dataset-specific statistics are computed exclusively from the training samples. For each channel $c \in$ {R,G,B}, the mean $\mu_c \in \mathbb{R}$ and standard deviation $\sigma_c \in \mathbb{R}$ are calculated as:

$$\mu_c = \frac{1}{N_{\text{train}}} \sum_{i=1}^{N_{\text{train}}} \frac{1}{H \cdot W} \sum_{h,w} X_i^{(h,w,c)}, \quad (35)$$

$$\sigma_c = \sqrt{\frac{1}{N_{\text{train}}} \sum_{i=1}^{N_{\text{train}}} \frac{1}{H \cdot W} \sum_{h,w} \left(X_i^{(h,w,c)} - \mu_c\right)^2}. \quad (36)$$

Each image is then normalized per channel:

$$\tilde{X}_c^{(h,w)} = \frac{X_c^{(h,w)} - \mu_c}{\sigma_c}, c \in \{\text{R,G,B}\}. \quad (37)$$

This normalization ensures zero-mean, unit-variance inputs, which improves training convergence and overall performance.

*Training Procedure and Early Stopping:* The model is trained using cross-entropy loss with label smoothing $\alpha = 0.1$:

TABLE III
PER-DEVICE PACKET STATISTICS

| Dev | Name | MAC | µ | σ |
|---|---|---|---|---|
| 0 | Dropcam | 30:8c:fb:2f:e4:b2 | 113.1 | 50.2 |
| 1 | HP Printer | 70:5a:0f:e4:9b:c0 | 151.7 | 237.9 |
| 2 | Triby Speaker | 18:b7:9e:02:20:44 | 143.7 | 222.5 |
| 3 | Light Bulbs LiFX Smart Bulb | d0:73:d5:01:83:08 | 109.8 | 79.9 |
| 4 | Netatmo Weather Station | 70:ee:50:03:b8:ac | 258.1 | 128.9 |
| 5 | Nest Dropcam | 30:8c:fb:b6:ea:45 | 1057.6 | 663.3 |
| 6 | TP-Link Day Night Camera | f4:f2:6d:93:51:f1 | 106.8 | 75.6 |
| 7 | Withings Aura Sleep Sensor | 00:24:e4:20:28:c6 | 167.7 | 138.6 |
| 8 | Netatmo Welcome | 70:ee:50:18:34:43 | 499.0 | 645.9 |
| 9 | Belkin Wemo Switch | ec:1a:59:79:f4:89 | 277.1 | 424.7 |
| 10 | Amazon Echo | 44:65:0d:56:cc:d3 | 124.3 | 168.2 |
| 11 | Samsung Galaxy Tab | 08:21:ef:3b:fc:e3 | 211.5 | 333.0 |
| 12 | Samsung SmartCam | 00:16:6c:ab:6b:88 | 293.1 | 239.0 |
| 13 | MacBook | ac:bc:32:d4:6f:2f | 519.2 | 613.6 |

$$L_{\text{CE}} = -\sum_{c=1}^{|D|} \tilde{y}_c \log(\hat{y}_c)\,, \tag{38}$$

where $\tilde{y}_c = (1-\alpha)\delta(c,y) + \alpha/|\,D\,|$ is the smoothed label distribution, and $\delta(c,y) = 1$ if $c = y$, otherwise $\delta(c,y) = 0$. The training configuration includes batch size $B = 16$, maximum epochs $E_{\max} = 50$, gradient clipping where the gradient norm $\|\nabla_\theta L\|_2$ is clipped to a maximum of 1.0 to prevent exploding gradients during backpropagation and early stopping. For each epoch $e \in \{1, \ldots, E_{\max}\}$ training loss and accuracy are computed as:

$$L_{\text{train/val}}^{(e)} = \frac{1}{N_{\text{train/val}}} \sum_{(X,y)\in\mathcal{D}_{\text{train/val}}} L_{\text{CE}}\big(\text{softmax}\big(f_\theta(X)\big), \tilde{y}\big)\,, \tag{39}$$

$$\text{Acc}_{\text{train/val}}^{(e)} = \frac{1}{N_{\text{train/val}}} \sum_{(X,y)\in\mathcal{D}_{\text{train/val}}} \mathbf{1}[\arg\max f_\theta(X) = y]\,. \tag{40}$$

When measuring validation loss $y$ is used instead of $\tilde{y}$ in (39) and vice versa for accuracy in (40) (without gradient computation). Training terminates early if $\text{Acc}_{\text{val}}^{(e)}$ does not improve for $P = 15$ consecutive epochs. This configuration balances computational efficiency, convergence speed, and generalization performance across the three ViT architectures.

## V. RESULTS AND DISCUSSION

This section reports the experimental findings obtained from the evaluation. We analyze traffic characteristics, assess DL performance and examine robustness of the system.

### *A. Experimental Configuration*

The experimental design explores the factorial combination of DL models, transformation method, resolution, segment length, and overlap fraction. Formally:

$$|\,\mathcal{C}\,| = 3_{(\text{ViT})} \times 2_{(\text{STFT/CWT})} \times 3_R \times 2_{(L_{\text{seg}})} \times 2_{(p_{\text{overlap}})}, \tag{41}$$

which include 3 ViT models of DeiT-Base, ViT-Small, and EfficientViT, 2 transformation methods of STFT and CWT, 3 time-frequency resolutions: R $\in$ {16, 32, 64}, 2 segment lengths $L_{\text{seg}} \in \{100, 500\}$ packets and 2 overlap levels $p_{\text{overlap}} \in \{0\%, 50\%\}$, a total of 72 configurations. This comprehensive factorial ablation design systematically evaluates how each hyperparameter and model architecture affects device identification accuracy and robustness.

TABLE IV
TRAIN, VALIDATION AND TEST SPLIT PER DEVICE

| $L_{\text{seg}}$ | $p_{\text{overlap}}$ | R | Train Set | | Val set | | Test set | |
|---|---|---|---|---|---|---|---|---|
| | | | Pkts | Images | Pkts | Images | Pkts | Images |
| 100 | 0% | {16,32,64} | 40000 | 400 | 8000 | 80 | 8000 | 80 |
| | 50% | | 20000 | 400 | 4000 | 80 | 4000 | 80 |
| 500 | 0% | | 75000 | 150 | 15000 | 30 | 15000 | 30 |
| | 50% | | 37500 | 150 | 7500 | 30 | 7500 | 30 |

*Dataset:* We used the UNSW IoT Traffic Dataset [17] containing packet traces from 14 diverse IoT devices. Our analysis used only packet-length of encrypted payloads to demonstrate a minimal-information attack scenario (even the packet directions are discarded intentionally). No payload content, decrypted fields, or protocol-specific headers were accessed or used at any point in the pipeline. As the objective of this work is to analyze packet length characteristics of encrypted wireless traffic across heterogeneous devices, the selection of devices (listed in Table III) was made to ensure diversity and sufficient data availability. To enable reliable development and evaluation, the dataset was systematically divided into training, validation, and testing subsets across the experimental configurations, as summarized in Table IV. The raw packet sequences for each device were partitioned into disjoint training, validation, and test pools before any segmentation was applied. Overlapping ($p_{overlap}$) windows were generated independently within each pool after partitioning, never across pool boundaries. This ensures that no packet from a test-pool sequence contributes to any training segment, regardless of the overlap configuration. The OOD evaluation used the same pre-partitioned device sequences under different parameters not seen during training.

### *B. Traffic Analysis*

*Packet Statistics:* The 14 IoT and consumer devices in the dataset demonstrate significant heterogeneity in packet size distributions, attributable to their diverse communication protocols and distinct traffic behaviors that form the basis for device-specific fingerprinting. Table III and Fig. 2 summarize the packet-length statistics of each device which were generated from over 20,000 packets per device, with statistical descriptors such as mean ($\mu$), standard deviation ($\sigma$) and coefficient of variation ($cv$) computed to capture both the central tendency and variability of transmitted packet lengths. Devices such as dropcam, LiFX smart bulb, Netatmo weather station, Nest dropcam, TP-Link camera, Withings sleep sensor and Samsung SmartCam exhibit relatively low variability ($cv \leq 0.8$), reflecting stable and periodic traffic patterns typical of sensing or monitoring applications. In contrast, devices like the HP Printer, Triby Speaker, Netatmo Welcome, Belkin Wemo Switch, Amazon Echo, Samsung Galaxy Tab, and MacBook show high variability ($cv \geq 1.0$), suggesting bursty or mixed interactive traffic driven by user activity, multimedia streaming, or dynamic network exchanges. The Nest Dropcam stands out with the highest mean packet length (≈1058 bytes) and high variability. All devices show minimum packet sizes around 43–66 bytes, corresponding to wireless control or acknowledgment frames and maximum values near 1500 bytes, reflecting the upper bound of standard IP-layer packet sizes in wireless networks. This inter-device heterogeneity in packet-size dynamics demonstrates that packet-length statistics encode robust, discriminative, and non-payload-dependent device signatures suitable for accurate IoT device fingerprinting.

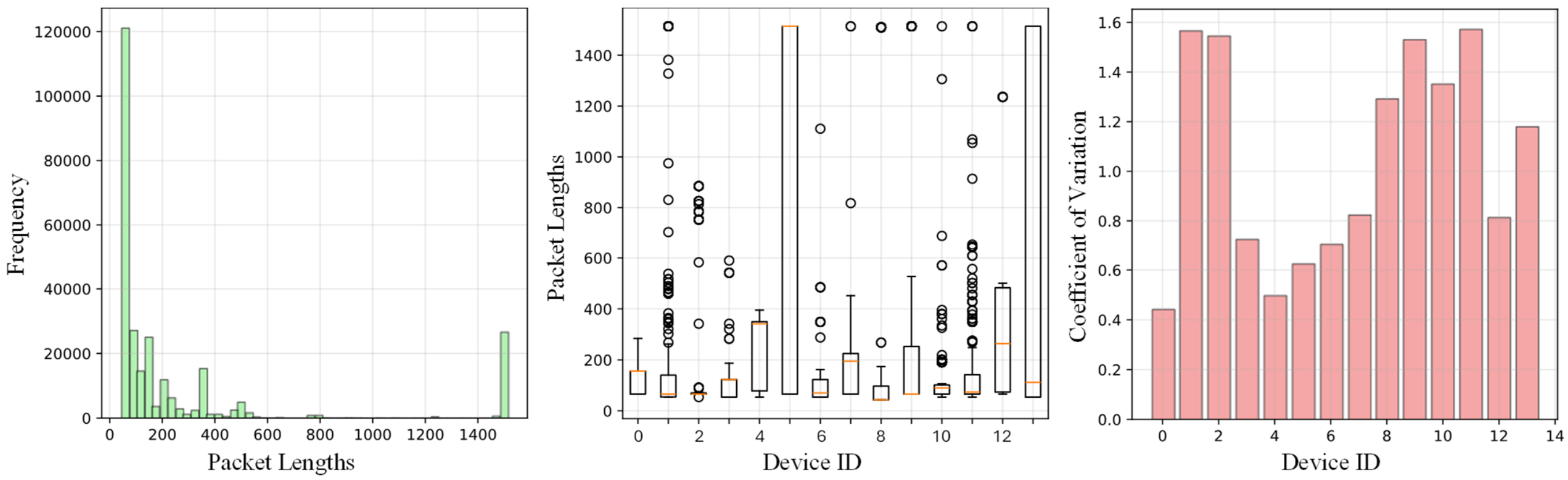

**Fig. 2.** Packet statistics of the devices in the dataset.

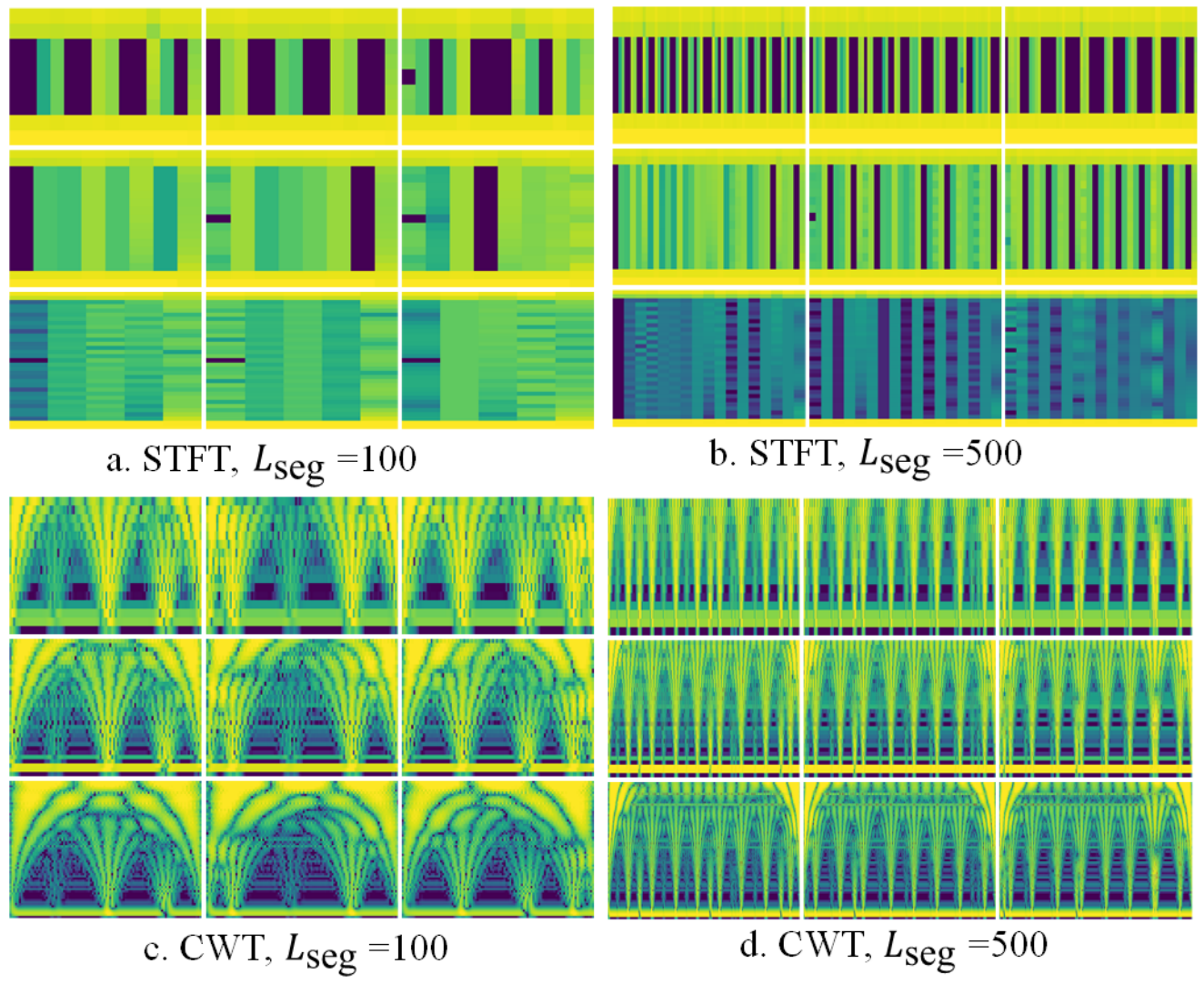

**Fig. 3.** Sample time-frequency representation of Device (0). The resolutions $R \in \{16,32,64\}$ top to bottom, respectively.

*Time-frequency Visual Inspection:* Fig. 3 illustrates a sample time-frequency representation sequence of device (0) (dropcam). For STFT, at coarse resolution ($R = 16$), both $L_{seg} = 100$ and $L_{seg} = 500$ spectrograms exhibit dominant vertical color bands encoding periodic energy in specific frequency bins correspond to short-term periodic bursts in packet transmission. At intermediate resolution ($R = 32$), the vertical bands subdivide into finer striations, and color gradations from dark teal to bright green become visible, representing packet-length variations around the mean. As resolution increases ($R = 64$), these structures become smoother and more continuous, capturing finer temporal variations and subtle frequency modulations within the shorter observation window. In contrast, spectrograms generated with $L_{seg} = 500$ reveal markedly richer temporal dynamics and denser vertical striping, reflecting the extended observation window's ability to capture more long-term fluctuations in packet-size behavior. The higher segment length also increases time-frequency density, allowing the persistence and periodicity of transmission bursts to emerge more clearly, particularly at lower resolutions ($R = 16, 32$), while the highest resolution ($R = 64$) captures both fine-grained temporal variability and background continuity.

CWT scalograms exhibit scale-dependent wavelet magnitude ridges and localized high-energy bands that capture device-specific temporal structure. The $L_{seg} = 100$ scalograms display pronounced patterns, while $L_{seg} = 500$ scalograms show these patterns extended across the full temporal span, indicating that wavelet scales are dynamically distributed across the longer duration. As resolution increases from $R = 16$ to $R = 64$, the wavelet coefficients become progressively finer, revealing more detailed oscillatory contours and multi-scale harmonics that highlight device-specific periodicities and transient bursts. The patterns arc-like subdivide into nested sub-patterns with fine striations interleaved within the gradients, capturing scale-dependent time-frequency features at multiple resolutions simultaneously. $L_{seg} = 100$ and $L_{seg} = 500$ scalograms exhibit visually distinct patterns, $L_{seg} = 500$ displays finer, more densely packed striations, while $L_{seg} = 100$ shows coarser sub-patterns concentrated in narrower frequency bands. At the finest resolution ($R = 64$), rich multi-scale texture dominates, where $L_{seg} = 100$ concentrates detail in mid-frequency regions (consistent with the device's stable 109-byte median), while $L_{seg} = 500$ extends energy across the full frequency range including ultra-low frequencies. This capture extended low-frequency periodicities spanning hundreds of packets that cannot be represented in brief 100-packet windows. Overall, increasing $L_{seg}$ expands temporal coverage, while increasing $R$ refines frequency precision as CWT can encode both transient and sustained traffic behaviors.

### *C. Empirical Validation of Time-frequency Representations*

To demonstrate that time-frequency representations encode more discriminative device structure than grayscale packet-image baselines, we conduct a controlled evaluation using a simple, architecture-agnostic pipeline. For grayscale conversion, packet-length sequences from each device are flattened into 1D vectors and split into non-overlapping windows of 100 and 500 packets. Each window is normalized to pixel intensities using device-specific min-max scaling, reshaped into a 2D matrix approximating a square aspect ratio, and resized via nearest-neighbor interpolation. The packet length-to-grayscale image conversion techniques is adapted from study [6], [7], [8], [9]. This evaluation is conducted independently of the ViT training pipeline in Section V.D; no ViT weights, task-specific features, or fine-tuned representations are used. The pipeline is deliberately simple and architecture-agnostic so that observed differences in clusterability and label recovery are attributable entirely to the input representation rather than to model capacity.

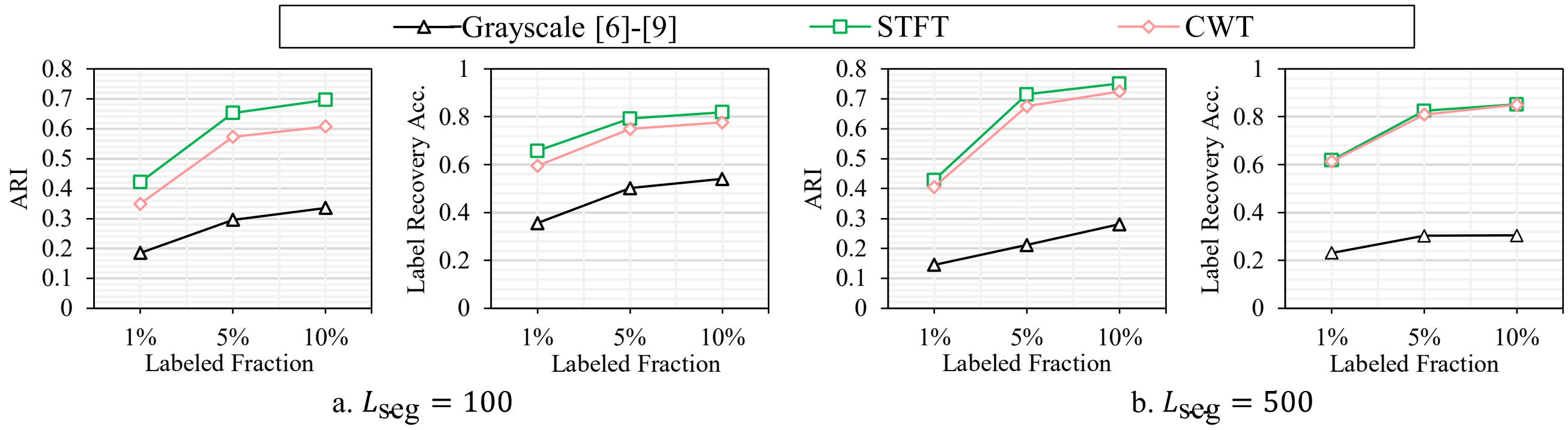

**Fig. 4.** Semi-supervised label propagation efficiency (LabelSpreading).

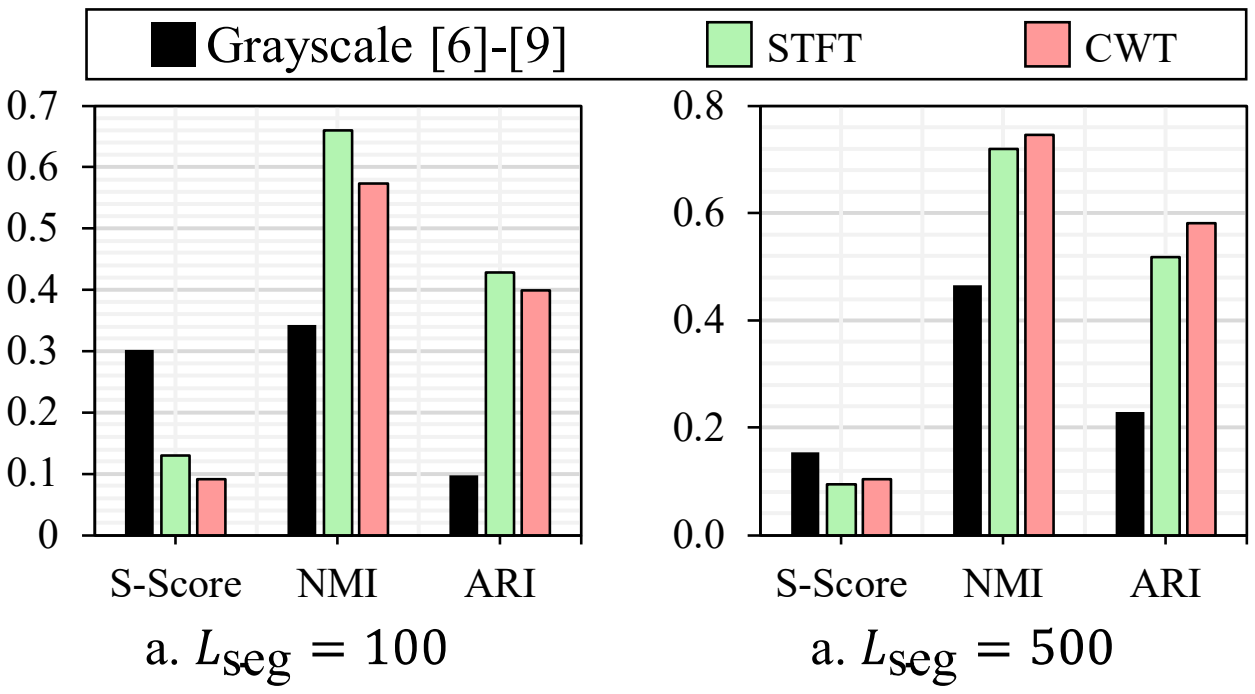

**Fig. 5.** Unsupervised KMeans clustering performance.

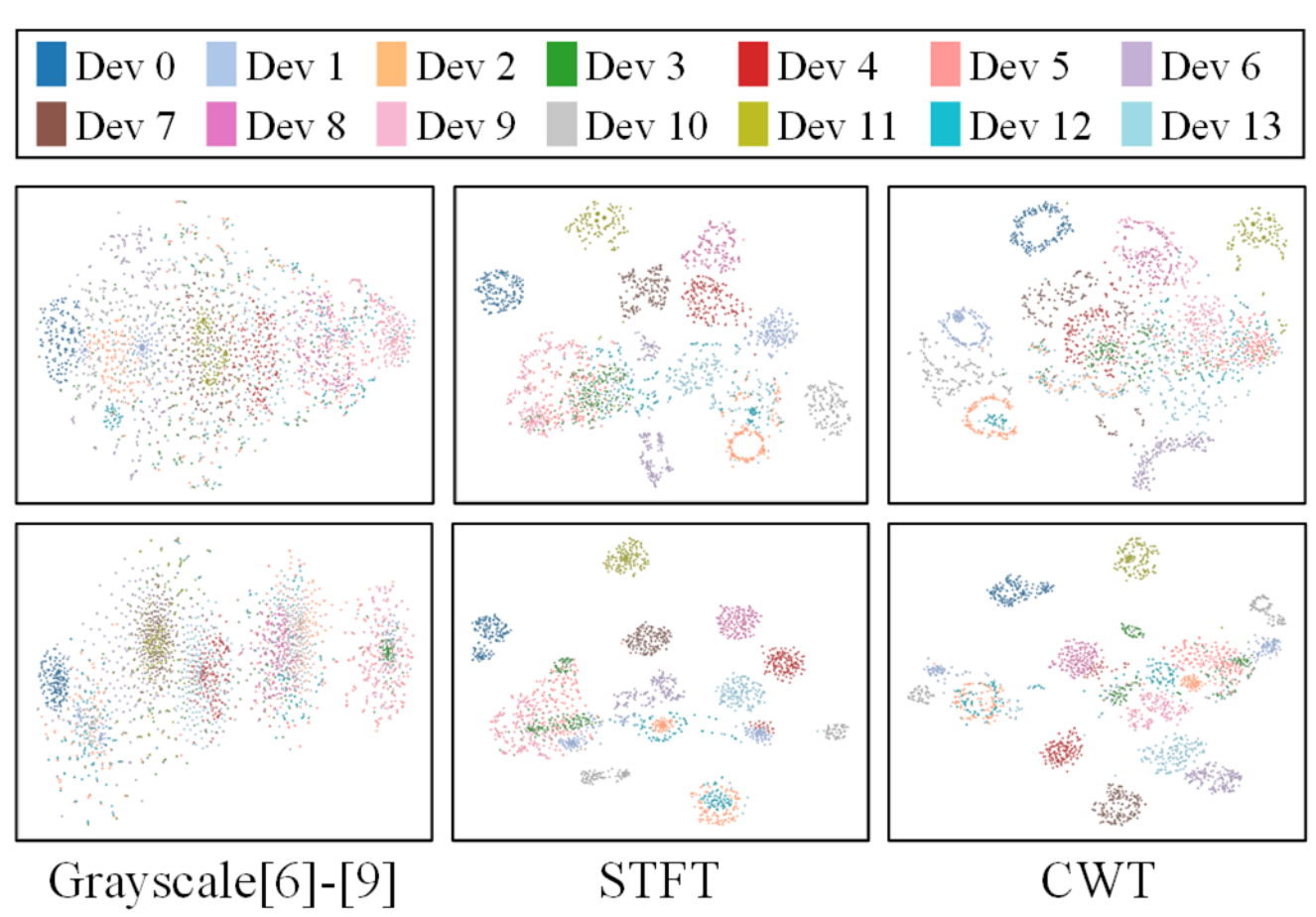

**Fig. 6.** t-SNE visualization of PCA-reduced representations of $L_{seg} = 100$ (top row) and $L_{seg} = 500$ (bottom row).

For evaluation, each image is flattened into a feature vector, standardized using StandardScaler, and projected via principal component analysis (PCA) retaining 95% explained variance. No deep feature extraction or task-specific modeling is performed. The pipeline is identical across all representations, and the only variable is the input encoding of grayscale, STFT ($R = 16$), or CWT ($R = 64$). The goal is to quantify the intrinsic clusterability and label-propagation properties of the representations using device labels only for post-hoc evaluation, not for model training.

*Semi-Supervised Label Recovery:* To assess label-propagation efficiency, we applied LabelSpreading, a transductive graph-based method that constructs a k-NN similarity graph over all samples and propagates labels through this graph. LabelSpreading directly probes representation geometry: successful propagation requires same-class samples to form cohesive neighborhoods in the feature space. Performance is evaluated exclusively on the initially unlabeled subset to ensure labeled samples do not inflate reported metrics. LabelSpreading is preferred over inductive self-training alternatives because its graph propagation provides a cleaner measure of intrinsic manifold structure, independent of classifier capacity. From Fig. 4, at 1% labeled data, STFT achieved +128% higher adjusted rand index (ARI) and +85% higher label recovery accuracy than grayscale at $L_{seg} = 100$, and +196% higher ARI and +168% higher label recovery accuracy at $L_{seg} = 500$. CWT achieved comparable gains of +89% ARI and +67% label recovery accuracy at $L_{seg} = 100$, and +179% ARI and +166% accuracy at $L_{seg} = 500$. These improvements were consistent across all evaluated label fractions (1%, 5%, 10%) and both segment lengths, demonstrating that time-frequency representations enable more effective label propagation than grayscale encodings even under minimal supervision.

*Unsupervised Clusterability:* KMeans clustering is applied to PCA-reduced features with the number of clusters fixed to the true number of device classes (K=14). Clustering proceeds without access to labels. Post-hoc, we compute adjusted ARI and normalized mutual information (NMI) as external evaluation metrics to measure alignment between discovered clusters and ground-truth device labels, alongside Silhouette score (S-Score) as an internal measure of geometric compactness independent of labels. STFT and CWT substantially outperform grayscale across both ARI and NMI for both segment lengths. As shown in Fig. 5, STFT improved

TABLE V
EVALUATION OF DL ACROSS ALL EXPERIMENTAL CONFIGURATIONS

| $L_{seg}$ | $p_{overlap}$ | $R$ | Model | STFT | | | | | CWT | | | | |
|---|---|---|---|---|---|---|---|---|---|---|---|---|---|
| | | | | Train (%) | Val (%) | Test (%) | F1 | CI width | Train (%) | Val (%) | Test (%) | F1 | CI width |
| 100 | 0% | 16 | DeiT-Base | 99.89 | 99.29 | 98.57 | 0.9857 | 1.25 | 100 | 99.55 | 97.23 | 0.972 | 1.88 |
| | | | ViT-Small | 98.82 | 97.41 | 92.41 | 0.914 | 2.95 | 99.98 | 99.73 | 97.86 | 0.9784 | 1.61 |
| | | | EfficientViT | 99.25 | 98.57 | 97.68 | 0.9768 | 1.7 | 99.93 | 99.64 | 96.61 | 0.9649 | 2.14 |
| | | 32 | DeiT-Base | 99.91 | 98.75 | 98.12 | 0.981 | 1.61 | 99.73 | 94.64 | 95.18 | 0.9472 | 2.42 |
| | | | ViT-Small | 99.88 | 98.84 | 98.39 | 0.9838 | 1.39 | 99.84 | 99.64 | 98.48 | 0.9847 | 1.43 |
| | | | EfficientViT | 99.96 | 98.75 | 97.41 | 0.9733 | 1.79 | 99.96 | 99.73 | 98.66 | 0.9865 | 1.34 |
| | | 64 | DeiT-Base | 100 | 96.61 | 93.93 | 0.9384 | 2.77 | 99.91 | 99.46 | 99.46 | 0.9946 | 0.8 |
| | | | ViT-Small | 99.95 | 96.79 | 93.21 | 0.9285 | 2.86 | 100 | 99.73 | 99.29 | 0.9928 | 0.8 |
| | | | EfficientViT | 99.98 | 96.79 | 92.59 | 0.9237 | 3.12 | 99.91 | 99.82 | 99.02 | 0.99 | 1.16 |
| | 50% | 16 | DeiT-Base | 99.93 | 98.84 | 98.84 | 0.9884 | 1.21 | 99.98 | 99.73 | 97.41 | 0.9736 | 1.79 |
| | | | ViT-Small | 99.91 | 98.84 | 98.21 | 0.982 | 1.61 | 99.98 | 99.64 | 97.5 | 0.9746 | 1.7 |
| | | | EfficientViT | 99.98 | 99.11 | 98.39 | 0.9839 | 1.34 | 99.98 | 99.91 | 96.52 | 0.9636 | 2.05 |
| | | 32 | DeiT-Base | 99.93 | 98.57 | 95.98 | 0.956 | 2.28 | 99.98 | 99.82 | 99.55 | 0.9955 | 0.8 |
| | | | ViT-Small | 99.88 | 98.66 | 97.32 | 0.9723 | 1.92 | 99.91 | 99.73 | 99.73 | 0.9973 | 0.54 |
| | | | EfficientViT | 99.96 | 98.57 | 97.68 | 0.9762 | 1.7 | 99.96 | 99.91 | 99.91 | 0.9991 | 0.27 |
| | | 64 | DeiT-Base | 99.98 | 96.25 | 94.64 | 0.946 | 2.59 | 100 | 99.73 | 99.73 | 0.9973 | 0.62 |
| | | | ViT-Small | 99.95 | 96.96 | 95.18 | 0.9511 | 2.5 | 99.98 | 99.55 | 99.29 | 0.9928 | 1.07 |
| | | | EfficientViT | 99.93 | 97.05 | 95.27 | 0.9513 | 2.46 | 99.95 | 99.91 | 99.46 | 0.9946 | 0.89 |
| 500 | 0% | 16 | DeiT-Base | 100 | 100 | 99.52 | 0.9952 | 1.19 | 100 | 98.81 | 97.14 | 0.9714 | 3.1 |
| | | | ViT-Small | 100 | 100 | 99.52 | 0.9952 | 1.19 | 99 | 98.57 | 97.62 | 0.9762 | 2.98 |
| | | | EfficientViT | 99.95 | 99.76 | 99.29 | 0.9928 | 1.67 | 99.95 | 99.29 | 94.05 | 0.9365 | 4.52 |
| | | 32 | DeiT-Base | 100 | 100 | 95.71 | 0.9971 | 3.81 | 100 | 99.76 | 100 | 1 | 0 |
| | | | ViT-Small | 99.9 | 99.52 | 95.48 | 0.9548 | 3.81 | 99.95 | 99.29 | 95 | 0.943 | 4.05 |
| | | | EfficientViT | 100 | 100 | 97.62 | 0.9758 | 3.1 | 100 | 100 | 99.29 | 0.9928 | 1.67 |
| | | 64 | DeiT-Base | 100 | 100 | 99.05 | 0.9905 | 1.79 | 100 | 100 | 99.29 | 0.9929 | 1.67 |
| | | | ViT-Small | 99.9 | 99.76 | 97.38 | 0.9734 | 3.1 | 100 | 100 | 99.29 | 0.9929 | 1.67 |
| | | | EfficientViT | 100 | 99.76 | 98.1 | 0.9809 | 2.62 | 99.81 | 100 | 100 | 1 | 0 |
| | 50% | 16 | DeiT-Base | 100 | 99.52 | 99.76 | 0.9976 | 0.71 | 100 | 98.81 | 98.57 | 0.9857 | 2.14 |
| | | | ViT-Small | 99.95 | 99.29 | 99.76 | 0.9976 | 0.71 | 99.86 | 98.57 | 98.81 | 0.988 | 1.9 |
| | | | EfficientViT | 100 | 100 | 99.76 | 0.9976 | 0.71 | 99.86 | 99.05 | 97.86 | 0.9785 | 2.62 |
| | | 32 | DeiT-Base | 99.95 | 99.52 | 96.52 | 0.9952 | 1.19 | 100 | 99.76 | 99.52 | 0.9952 | 1.19 |
| | | | ViT-Small | 99.95 | 99.52 | 99.05 | 0.9904 | 1.79 | 99.95 | 99.52 | 99.05 | 0.9905 | 1.67 |
| | | | EfficientViT | 99.95 | 99.52 | 99.52 | 0.9952 | 1.19 | 100 | 100 | 100 | 1 | 0 |
| | | 64 | DeiT-Base | 100 | 99.52 | 98.81 | 0.9877 | 2.14 | 100 | 100 | 98.57 | 0.9856 | 2.27 |
| | | | ViT-Small | 99.95 | 99.52 | 98.33 | 0.9832 | 2.62 | 100 | 100 | 100 | 1 | 0 |
| | | | EfficientViT | 99.95 | 99.76 | 98.81 | 0.9881 | 2.14 | 100 | 100 | 100 | 1 | 0 |

**Notation:** Green cells highlight the best test accuracy within each overlap and segment-length group.

ARI by +338% at $L_{seg} = 100$ and +125% at $L_{seg} = 500$ over grayscale, while CWT improved by +308% and +153% respectively. NMI improvements range from +57% to +93% across segment lengths. Grayscale yields higher Silhouette scores despite substantially lower ARI and NMI. This indicates that grayscale features form geometrically compact clusters in the PCA space that do not correspond to true device boundaries – the representation clusters on low-level byte patterns rather than device-discriminative structure. STFT and CWT sacrifice internal compactness for ground-truth alignment, which is the meaningful criterion for device classification.

Fig. 6 presents t-SNE projections of PCA-reduced features (200 samples per class, 2,800 total per representation) colored by ground-truth device class. t-SNE is applied to PCA-compressed features rather than raw pixel vectors, following standard practice to improve embedding quality and reduce computational cost. Identical t-SNE hyperparameters are applied across all representations for fair comparison. These visualizations provide qualitative support that STFT and CWT projections exhibit well-separated, spatially distinct class clusters, while grayscale projections show substantially greater inter-class overlap and less defined cluster boundaries.

*D. ViT Evaluation on Time-frequency Representations*

The primary goal of CONTEX-T is to evaluate how well DL architectures can exploit the information encoded in the spectrograms and scalograms for device identification. Three transformer-based models (DeiT-Base, ViT-Small, and EfficientViT) were systematically compared across the 72 configurations. Table V presents the model performance.

Across all configurations and model architectures, the models consistently demonstrated superior performance. For STFT-based spectrograms, test accuracy typically exceeded 98% (as shown by green-highlighted cells in the table). The best performance for short segments ($L_{seg} = 100, R = 16, p_{overlap} = 0\%$) was achieved by DeiT-Base, with 98.57% accuracy and an F1-score of 0.9857. Accuracy improved on longer segments, exceeding 99%. With $p_{overlap} = 50\%$, performance improved further for both segment lengths, achieving best accuracy with low resolution ($R = 16$).

TABLE VI

EVALUATION OF BEST MODELS ACROSS OOD SET FOR CROSS-CONFIGURATION EVALUATION

| Best Model | | | | Accuracy on OOD Evaluation Set (%) | | | | | | | | | | | |
|---|---|---|---|---|---|---|---|---|---|---|---|---|---|---|---|
| | | | | $L_{seg}$ =100 | | | | $L_{seg}$ =200 | | | | $L_{seg}$ =500 | | | |
| **Method** | $L_{seg}$ | ***R*** | **Best Trained Model (*p*)** | ***p*=0%** | ***p*=25%** | ***p*=50%** | ***p*=75%** | ***p*=0%** | ***p*=25%** | ***p*=50%** | ***p*=75%** | ***p*=0%** | ***p*=25%** | ***p*=50%** | ***p*=75%** |
| STFT | 100 | 16 | DeiT-Base (*p*=50%) | 98.57 | 98.33 | 98.81 | 98.33 | 100 | 100 | 99.76 | 99.76 | 97.38 | 96.9 | 99.52 | 100 |
| | | 32 | EfficientViT (*p*=50%) | 97.62 | 98.1 | 97.62 | 97.38 | 99.52 | 99.29 | 99.29 | 99.52 | 98.81 | 98.33 | 100 | 99.76 |
| | | 64 | ViT-Small (*p*=50%) | 95.24 | 94.29 | 94.76 | 93.1 | 82.86 | 82.86 | 82.62 | 81.67 | 77.86 | 77.14 | 78.33 | 78.1 |
| | 500 | 16 | DeiT-Base (*p*=50%) | 29.52 | 30 | 29.52 | 30.48 | 79.29 | 79.76 | 81.43 | 83.57 | 99.52 | 99.76 | 99.76 | 100 |
| | | 32 | ViT-Small (*p*=50%) | 62.86 | 63.57 | 63.33 | 65 | 85 | 85.71 | 86.67 | 90.24 | 99.29 | 99.52 | 99.52 | 99.52 |
| | | 64 | EfficientViT (*p*=50%) | 47.86 | 47.38 | 47.62 | 48.57 | 83.81 | 83.57 | 84.29 | 83.81 | 99.05 | 98.57 | 98.81 | 98.57 |
| CWT | 100 | 16 | Vit-Small (*p*=50%) | 98.33 | 97.38 | 97.14 | 97.38 | 69.52 | 71.9 | 70 | 70.24 | 22.62 | 23.57 | 23.57 | 24.76 |
| | | 32 | Vit-Small (*p*=50%) | 99.76 | 99.76 | 99.29 | 99.05 | 82.38 | 83.81 | 82.86 | 81.43 | 31.19 | 32.86 | 30.95 | 31.43 |
| | | 64 | DeiT-Base (*p*=50%) | 99.76 | 100 | 100 | 100 | 92.62 | 91.9 | 90.95 | 92.14 | 30.48 | 32.38 | 31.9 | 32.86 |
| | 500 | 16 | DeiT-Base (*p*=50%) | 30 | 28.33 | 30.24 | 29.05 | 56.19 | 56.67 | 53.33 | 56.19 | 97.86 | 97.14 | 98.57 | 97.86 |
| | | 32 | DeiT-Base (*p*=0%) | 51.9 | 51.9 | 54.05 | 50.71 | 64.76 | 64.29 | 64.05 | 66.43 | 100 | 100 | 100 | 100 |
| | | 64 | DeiT-Base (*p*=50%) | 47.86 | 50.71 | 48.57 | 47.62 | 79.52 | 77.62 | 78.1 | 79.29 | 98.57 | 98.33 | 98.57 | 98.1 |

**Notation**: $p$ denotes the $p_{\text{overlap}}$. The red border indicates test set with configuration that the best model was trained on, the rest are OOD set.

CWT-based scalograms showed comparable or slightly better results, especially at higher resolutions ($R = 32, 64$) and longer segments ($L_{seg} = 500$). The highest test accuracy was achieved by EfficientViT and ViT-Small for $L_{seg} = 500$ and $R = 32, 64$ with $p_{overlap} = 50\%$. Even for shorter segments ($L_{seg} = 100$), CWT models maintained high performance, with F1-scores up to 0.9946 and minimal variance.

*Impact of Segment Length, Resolution, and Overlap:* Increasing the segment length from 100 to 500 packets greatly enhanced performance stability, as longer segments capture richer temporal and spectral features that reveal device-specific periodicities and burst patterns. Both STFT and CWT achieved their best results at $L_{seg} = 500$. Lower resolution ($R = 16$) performed particularly well for STFT, while intermediate resolutions ($R = 32$) offered the best balance between accuracy and generalization. At the highest resolution ($R = 64$), CWT models sometimes achieved better performance, and STFT results slightly declined due to sensitivity to fine frequency variations. Additionally, applying a 50% overlap increased sample diversity and temporal context, consistently yielding modest improvements in accuracy and F1 by enhancing robustness to minor temporal shifts.

*Model Comparison:* Among the ViT variants, DeiT-Base generally achieved the highest performance across both STFT and CWT, particularly for moderate or longer $L_{seg}$ and higher $p_{overlap}$. EfficientViT also matched or outperformed DeiT-Base in CWT configurations, showing that lightweight transformer designs do not compromise accuracy when temporal and spectral features are used. ViT-Small achieved strong results but lagged DeiT-Base and EfficientViT in more challenging or lower-resolution scenarios, likely due to its reduced model capacity. Notably, in several CWT configurations ($L_{seg} = 500, R = 32, 64, p_{overlap} = 50\%$), both EfficientViT and ViT-Small achieved high scores, indicating strong separability of device fingerprints when time-frequency information is captured at the optimal scale.

*Statistical Robustness and Confidence Analysis:* The confidence intervals reported across all configurations underscore the statistical robustness of the proposed approach. Out of 72 total configurations, 46 configurations achieved confidence interval widths below 2.0%, of which 15 configurations achieved below 1.0% confidence intervals. These confidence intervals demonstrate that the frequency-domain representations yield highly discriminative patterns with minimal stochastic variation. The configurations achieving lowest-width confidence intervals were concentrated in longer-$L_{seg}$, higher-$p_{overlap}$, and higher- $R$ settings, where the models had access to sufficiently rich temporal and spectral information to make accurate, unambiguous classifications. Few (8 configurations) exhibited wider confidence intervals of 3-4% corresponding to scenarios with reduced data availability or lower resolution. These wider intervals still have strong performance with accuracy over 95%.

### *E. Cross-Configuration Evaluation*

Evaluation of various configuration demonstrated excellent performance under matched conditions. To further assess model robustness, the best-performing model from each configuration was evaluated on held-out out-of-distribution (OOD) evaluation sets of three segment lengths ($L_{seg} \in \{100, 200, 500\}$) and four overlap fractions ($p_{overlap} \in \{0\%, 25\%, 50\%, 75\%\}$) which were not used for training, validation or test. This OOD set simulates scenarios where observation windows may vary in duration and $p_{overlap}$, enabling assessment of model generalization beyond matched conditions. Table VI shows the results of cross-configuration.

For STFT-trained models, those trained in shorter segments and lower $R$ generalize far better across all OOD conditions. Models trained with $L_{seg} = 100$ (at $R = 16, 32$) consistently achieve 98–99% accuracy across all $L_{seg}$ and $p_{overlap}$. It maintains or even improves performance when tested on longer unseen segments, achieving over 99% accuracy on $L_{seg} = 200, 500$, exceedingly even the 98.81% baseline from matched training conditions. This suggests that short-segment training compels models to focus on core, transferable device-specific features that generalize across temporal scales. In contrast, models trained on longer segments ($L_{seg} = 500$) perform poorly (29–48% accuracy) when tested on shorter windows, likely due to overfitting to long-term temporal patterns. Across all configurations, $p_{overlap}$ variation caused minimal degradation (less than 1-2% deviation), confirming that STFT-based models learn overlap-invariant, stable representations.

CWT models demonstrate strong consistency across $p_{overlap}$ conditions but high sensitivity to $L_{seg}$. When trained on $L_{seg} = 100$ with $R = 32, 64$, they achieve outstanding accuracy (99.76–99.94%) on matching $L_{seg}$ test sets with OOD $p_{overlap}$ but collapse to 22–33% accuracy on longer segments. Conversely, models trained on $L_{seg} = 500$ (e.g., $R = 32, p = 0\%$) reach exceptional accuracy across all $p_{overlap}$ settings for the same segment length yet drop to around 52% when tested on shorter segments. This pattern indicates that while CWT's multi-resolution nature enables precise modeling within a fixed temporal scale, it fails to generalize when the temporal context changes, as learned wavelet scales become misaligned with the new segment structure. Similar to STFT-trained models, $p_{overlap}$ variation caused minimal degradation (less than 1% standard deviation), confirming overlap-invariance in CWT.

Across all ViT models, those trained with $p_{overlap} = 50\%$ consistently demonstrated the best robustness and performance across OOD evaluation sets, except for one CWT model ($L_{seg} = 500, R = 32, p = 0\%$). This advantage arises from smoother temporal transitions that help models learn segment-length independent features. Notably, even though the same number of samples (400) are used for training, the $p_{overlap} = 50\%$ setting requires only half the raw packet data compared to the non-overlapping case (refer to Table IV) yet still achieves superior accuracy and stability. Overall, STFT models trained on short segments ($L_{seg} = 100$) and lower resolutions ($R = 16, 32$) show the strongest robustness, maintaining high accuracy across varying $L_{seg}$ and $p_{overlap}$ conditions. CWT models achieve superior accuracy only when $L_{seg}$ match; their performance deteriorates sharply otherwise.

### F. Discussion

While the reported accuracy (>99%) is comparable to prior studies [6], [7], [8], [9], the proposed framework provides important advantages by exposing temporal and spectral structure explicitly, supporting more interpretable analysis as discussed in Section V.C. The cross-configuration validation on OOD set discussed in Section V.E. also demonstrate robustness across different configurations. STFT employs uniform frequency bins (fixed resolution $\Delta f$) across all time points. This uniform structure enables robust cross-segment generalization because frequency bins remain valid regardless of $L_{seg}$, explaining why STFT-trained models on $L_{seg} = 100$ generalize well to unseen $L_{seg} = 200, 500$ OOD test windows. In contrast, CWT employs adaptive scale-dependent resolution where coarse scales capture low frequencies with poor temporal precision while fine scales capture high frequencies with excellent temporal precision, producing hierarchical wavelet decompositions visible as nested patterns in Fig. 3. This adaptive resolution enables CWT to achieve exceptional classification in many instances within matched $L_{seg}$ by optimally allocating resolution across the observed frequency-time space. However, the scale-dependent patterns become fundamentally misaligned when $L_{seg}$ change. The visual divergence between CWT $L_{seg} = 100$ and $L_{seg} = 500$ scalograms (densely packed high-resolution patterns in 100-packet windows versus extended low-frequency details in 500-packet windows) reflects learned scales that are incompatible across segment-length shifts. This scale-dependence explains CWT's superior in-distribution accuracy and severe cross-segment performance collapse. This is because learned wavelet scales perfectly encode device signatures within matched $L_{seg}$ lengths but become fundamentally misaligned when tested on different durations. Therefore, for practical implementation, STFT offers robustness to variable deployment conditions and superior cross-segment generalization, while CWT provides theoretically optimal feature extraction within constrained operational contexts where segment lengths are fixed and data volume permits fine-resolution multi-scale analysis.

*Limitations and Future Direction*: As an initial study of time-frequency as interpretable representations of packet-length sequences, CONTEX-T is evaluated on the UNSW IoT dataset to establish proof-of-concept discriminative power. The exceptional accuracies validate that spectral analysis exposes device-specific communication signatures, establishing the technique's potential with minimal-information (utilizing only packet-length and low number of packets). Evaluation across diverse wireless environments, different datasets, evolving traffic/device conditions, real-time multi-device scenarios remain important future work to quantify transferability and robustness limits. Furthermore, CONTEX-T reveals significant contextual privacy leakage in traffic metadata. The interpretability of time-frequency representations enables targeted defense development, while comprehensive countermeasure evaluations across diverse conditions remains important future work to quantify mitigation effectiveness.

## VI. Conclusion

This paper presents CONTEX-T, a novel framework demonstrating that encrypted IoT traffic remains vulnerable to device-identity inference through time-frequency analysis of packet-length metadata. By decomposition of raw packet-length sequences into rich temporal-spectral representations, the framework achieved exceptional device classification accuracy. The comprehensive factorial evaluation across multiple hyperparameter configurations revealed that packet-length distributions encode deterministic, device-specific signatures robust to strong encryption protocols. STFT-based models demonstrated superior cross-segment generalization, maintaining consistent high accuracy across variable observation windows, making them practical for real-world deployment. Conversely, CWT models achieved optimal performance within matched training conditions but exhibited segment-length sensitivity. While these findings highlight the need for privacy-preserving defenses that obfuscate metadata and traffic patterns, the development of such defenses represents crucial future work. Therefore, this work establishes an important foundation for future research into countermeasures against passive, undetectable privacy exploitations in IoT networks.